# Quick-connect scanning tunneling microscope head with nested piezoelectric coarse walkers

**Angela M. Coe, Guohong Li, and Eva Y. Andrei** [a)]

**AFFILIATIONS:**
Department of Physics and Astronomy, Rutgers, the State University of New Jersey, Piscataway, New Jersey 08854, USA

[a)]**Author to whom correspondence should be addressed:** eandrei@physics.rutgers.edu

**ABSTRACT:**
To meet changing research demands, new scanning tunneling microscope (STM) features must constantly evolve. We describe the design, development, and performance of a modular plug-in STM which is compact and stable. The STM head is equipped with a quick-connect socket that is matched to a universal connector plug, enabling it to be transferred between systems. This head can be introduced into a vacuum system via a load-lock and transferred to various sites equipped with the connector plug, permitting multi-site STM operation. Its design allows reliable operation in a variety of experimental conditions, including a broad temperature range, ultra-high vacuum, high magnetic fields, and closed-cycle pulse-tube cooling. The STM's compact size is achieved by a novel nested piezoelectric coarse walker design which allows for large orthogonal travel in the X, Y, and Z directions, ideal for a studying both bulk and thin film samples ranging in size from mm to μm. Its stability and noise tolerance is demonstrated by achieving atomic resolution in ambient conditions on a laboratory desktop with no vibrational or acoustic isolation.

## I. INTRODUCTION

Modern scanning probe microscopy (SPM) provides access to a wide range of material characteristics at the nanometer scale, including topographic, electrical, magnetic, mechanical, thermal, and optical properties[1-12] with applications across the fields of physics, chemistry, material science, life science, and engineering. Scanning tunneling microscopy (STM), invented by Binnig and Rohrer in 1981[13-15], was the first SPM technique to be developed. STM offers high resolution real-space imaging of the atomic-scale surface structure on conducting materials, exploration of the material's electronic properties through scanning tunneling spectroscopy (STS), and serves as a tool for building nanoscale structures. Consequently, STM has become an essential tool to study exotic quantum states in materials. Interest in quantum materials has expanded quickly, from mm-sized single crystals and epitaxial thin films to μm-sized 2D materials and Moiré structures. The advent of scientific advances requiring control of ever more complex environments, led to the development of new STM head features that incorporate diverse functionalities.

Early STMs explored sample topography and atomic arrangements. As interest in understanding electronic properties arose, STS resolution was improved by developing cryogenic STM systems[16-22]. The higher energy resolution of low temperature (LT) STMs opened new frontiers in the study of correlated electronic phases including superconductivity and charge density waves. LT-STMs could be placed in superconducting magnets, thereby providing access to superconducting vortex phases[1,23] and Landau Level spectroscopy[24-26]. As exploration turned to surface sensitive materials[27-29], STMs were developed compatible with high vacuum (HV) and ultra-high vacuum (UHV) systems[30,31]. Sample size and type studied has impacted the STM's coarse motor requirements and electrical needs. Early samples were mm-sized, only requiring Z coarse motion to approach a tip to the sample, whereas the recent focus on 2D μm-sized samples requires X, Y, and Z coarse motion to navigate the STM tip to the sample area as well as electrical isolation to allow electrostatic gating.

Despite remarkable advances in STM technology, most standard STMs remain mounted to a fixed location via a rigid mechanical joint and a strong electrical connection that does not permit easy removal. The requirement for precise sample and tip alignment can be challenging for studies of 2D μm-sized samples. For example, within an LT-STM located in a superconducting magnet where optical access is unavailable, aligning the tip to a typical 2D sample such as graphene[32] or transition metal dichalcogenide[33-36], requires techniques for finding the sample without crashing the tip. To overcome these drawbacks, it is desirable to have a detachable STM that can be transferred between different sites. This capability facilitates sample and tip exchange, enables precise tip-sample alignment with optical access, and provides access to various experimental conditions. Furthermore, it allows for head exchange, creating the opportunity for more than one type of SPM head to be operated within a system.

Modern fixed STMs require high stability, complex wiring, large coarse motor travel in X, Y, and Z directions, and compatibility requirements for cryogenic operation, vacuum technology, and magnetic fields. To these inherent fixed STM requirements, the detachable feature adds additional requirements for a pluggable electrical connection, attachments for transferability within vacuum systems, and a compact size for ease of transfer and stability. Detachable STMs[37-40] are beginning to emerge in dilution refrigerator systems, but their applications have yet to be fully realized.

Here we describe the design and construction of an STM head[41] featuring an electrical quick-connect, vacuum transferability, and an ultra-compact nested coarse walker module. The quick-connect feature replaces the STM electrical wiring with a pluggable connection, making it easy to transfer the STM between different sites. The quick-connect socket is part of the STM head while the connector plug can be mounted in any site were STM operation is desired. This enables the STM head to be transferred between various sites within a system and also between different systems. The use of a nested piezoelectric coarse walker design allows a significant reduction of the STM head size resulting in higher stability and improved portability. This reduced size is achieved by nesting the X, Y, and Z walkers inside each other while maintaining large orthogonal coarse travel. The stability of this modular STM was demonstrated by achieving atomic resolution on highly oriented pyrolytic graphite (HOPG) in ambient conditions on a desktop with no vibration isolation.

## II. GENERAL OVERVIEW & CONCEPT

The STM head [Fig. 1] is 55.5mm tall with its largest diameter 36mm. Materials used in its construction were selected based on mechanical, thermal, and vacuum properties. Electrical connections are made via the quick-connect socket located at the bottom of the STM head. The mid-section of the STM contains the nested coarse walker module, which allows large scale tip movement for exploring μm-sized sample areas. A piezoelectric scanner mounted to the nested coarse walker allows fine control of the tip motion for atomic resolution within tunneling range of the sample. Tips are installed in a flag-based holder and plugged into the top of the piezoelectric scanner. Samples are mounted to a flag holder and plugged into the top frame of the STM head. Three types of transfer attachments are incorporated into the STM head to make it transferable within a vacuum system, they are a puck, flag, and a top plate.

To illustrate the concept and benefits of a detachable STM, we outline in [Fig. 2] several applications that have been successfully implemented. The STM head requires only the electrical connector plug to enable its operation at any particular station. We installed a connector plug in a desktop setup operating in ambient conditions which allows for easy initial testing of samples and tips before the STM head is transferred to a vacuum system. Subsequently the STM head can be introduced into a vacuum system through a load-lock and transferred to different stations using vacuum manipulators. A second connector plug was mounted within a UHV room temperature (RT) chamber with optical access which is convenient for installing tips and samples into the STM head, alignment of tips to a μm-sized sample, and for tests of sample integrity. A third connector plug, within the same UHV system was mounted into the bore of a superconducting magnet installed in a closed-cycle cryostat with no optical access. This provides STM access to low temperatures (down to ~ 4K) and magnetic fields (up to 9T). Another connector plug was mounted in a $^3$He system, where the STM is cooled to temperatures down to 300mK.

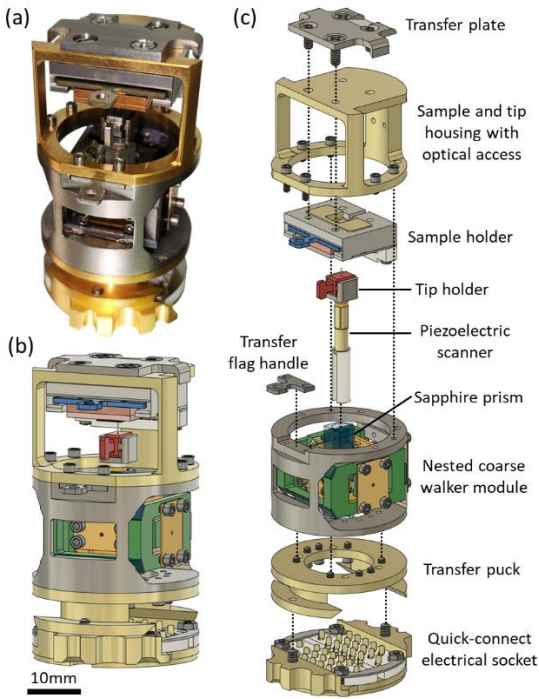

**FIG. 1:** Photograph and schematic of the STM head. (a) Front-view photograph of STM head. (b) Schematic of STM head. (c) Exploded-view schematic of the STM with components labeled. Electrical wiring in the schematics is omitted for clarity.

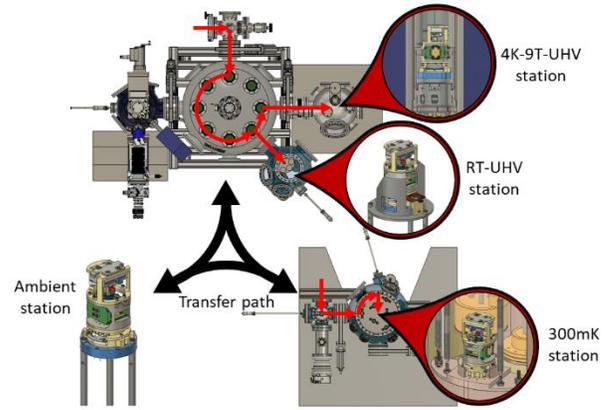

**FIG. 2:** Example transfer path of STM head through various systems to different stations equipped with the connector plug. Black arrows show the STM head transfer path between systems and red arrows show the transfer path within a system. Image at bottom left corner shows the STM head plugged into an electrical connector plug in ambient conditions on a desktop. Top image depicts a cryogen-free closed-cycle cryostat UHV system equipped with two connector plugs. One is mounted at RT for sample and tip exchange, alignment, and testing. The other connector plug is located within a superconducting magnet situated in a closed-cycle cryostat with temperatures near 4K and magnet fields up to 9T. This connector is used as an electrical vacuum feedthrough and is mounted on a vibration isolation unit. Bottom image on the right depicts a $^3$He system with temperatures down to 300mK where the STM head is joined with a connector plug with vibration damping.

## III. QUICK-CONNECT DESIGN AND CONSTRUCTION

STM heads are challenging to handle, partially due to competing wiring requirements, including separate high voltage, low voltage, and coaxial lines. To make a detachable head, a pluggable electrical connector must be designed to accommodate the various needs. The industry standard for sturdy and reliable connectors is the military circular connector which utilizes a socket-pin electrical connection locked together with a threaded outer ring. This design however is not practical for a pluggable STM connector which must be manipulated within a vacuum system, where handling ability is limited. Since STMs operate in a quiet and stable environment, a weaker connector with minimal locking needs can be utilized, such as a spring-pin interaction. The STM described here uses a quick-connect comprised of a socket with 34 spring contacts [Fig. 3] that are fixed to the bottom of the STM and of a matching connector plug with pin contacts.

The socket spring contacts (gold-plated beryllium copper, BeCu)[42] are fixed with insulating epoxy[43] to an insulating Macor insert which is clamped (304 stainless steel, SS) to the STM head frame (gold-plated copper 101), [Fig. 3(a)]. It is important to make the spring-pin interaction force adjustable and balanced. If the force is too large, the STM will be too difficult to plug in. If too small, the electrical connection will be unreliable. To balance the cumulative spring-pin force, the socket connector is split into two parts. The springs on the two parts are oriented in opposite directions which balances the spring force and centers the STM when inserted onto the connector plug. The position of the springs in the socket connector is made adjustable by loosening the clamps, allowing the spring-pin force to be modified. Here the springs have been positioned such that a slight push on top of the STM head, equivalent to about 300 grams-force or about 0.29N, is needed to fully insert the STM head (92 gr) into the connector plug. The socket connector has proven to be highly reliable with no noticeable deterioration after hundreds of connection events.

The connector plug[44] is comprised of pin contacts (nickel-plated 330 SS) that are fixed with insulating BPS glass within a 304 SS support frame, [Fig. 3(b)]. These pins were chosen for their reliability and durability, a necessary requirement for plugs mounted in hard to access locations. The pin diameter matches that of industry standard UHV compatible pins, 0.76mm. Running along the outer diameter of the connector are holes used for mounting or as alignment guides between the STM head socket and plug used during insertion. The connector plug is air-tight and can be used as a sealed vacuum feedthrough.

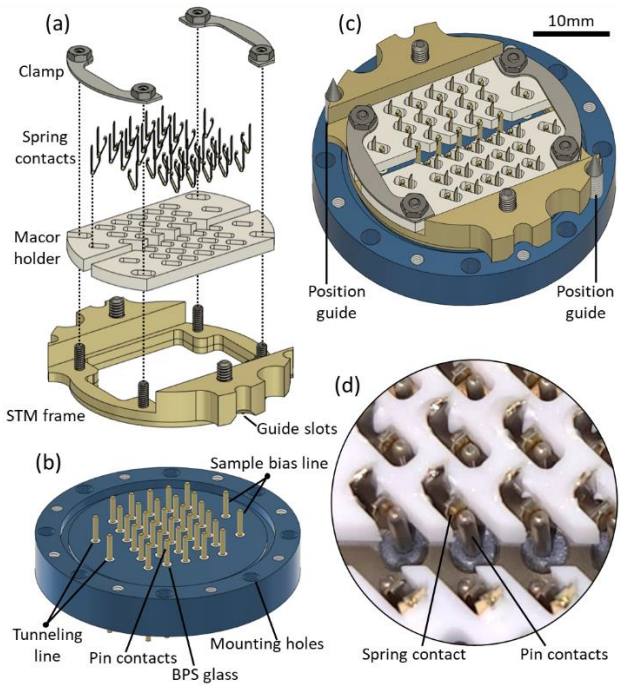

**FIG. 3:** STM head quick-connect electrical socket and matching connector plug. (a) Exploded schematic of quick-connect socket located at the bottom of the STM head. Thirty-four electrical spring contacts are held in a Macor insulator frame. The Macor frame is fixed to the STM head copper framing with two metal clamps. (b) Schematic of the connector plug showing the pin contacts and holes running along the outer diameter for mounting and position guide attachments. (c) Schematic of the STM base connector plugged into the pin connector. (d) Photograph of the spring-pin interaction.

When inserting the STM head, position guides aid the springs in locating the pins, [Fig. 3(c)]. These guides can take many forms, as illustrated in [Fig. 2]. An example of such a position guide is having two pointed screws fixed to the mounting holes in the pin connector. With a height greater than that of the pins, the screws interact with the guided slots in the STM head first as the STM is lowered to the connector plug. These screws serve to align the STM head to the connector and allow the springs to plug into the pins. The spring-pin interaction of a fully plugged in STM head connector is shown in [Fig. 3(d)]. The STM head socket is shaped to maximize surface contact to the connector plug to provide a strong thermal link at low temperatures. The socket is made from copper, for its high thermal conductivity, and gold-plated to prevent oxidation.

Since the connector is not coaxial, there can be measurable cross-talk between neighboring electrical contacts if the wire arrangement is not carefully considered. Electrical cross-talk was reduced by physically separating the tunneling and sample bias lines from each other and all other voltage signals. The connector is arranged with a central grouping of 30 contacts and four isolated contacts, two on each side of the central grouping [Fig 3(b)]. The tunneling and sample bias coax[45] lines were placed on either side of the central grouping on the isolated contacts. The high and low voltage twisted-pair[46] lines of the coarse module, piezoelectric scanner, additional sample connections, and temperature sensors were arranged on the central 30 contacts. The ground lines were placed between each section for best isolation. To simplify wiring within the STM head, holes were created in the tops of the socket spring contacts through which the electrical wires were threaded and anchored in place with conductive epoxy[47].

## IV. TRANSFERABILITY WITHIN VACUUM SYSTEMS

In order to be transferable within a vacuum system the STM head must have mechanical attachments that can engage with vacuum manipulators. To utilize commercial vacuum manipulators, the attachments must follow industry standards. Attachment types vary depending on operation requirements, such as travel distance, ease of attachment and detachment, optical access availability, and complexity of performed action. Three methods of transfer have been implemented in this STM head design.

Puck-styles are the simplest vacuum attachments. They are large structures that have a simple engagement method. They tend to be used for long range travel of objects between vacuum chambers, as indicated by the transfer path of the red arrow in [Fig. 2]. Transferable objects with a puck-style attachment have a circular horizontal groove into which manipulators fit with a u-shaped fork end[48,49]. The weight of the puck sits on the fork with gravity holding it in place during transfer. Our STM head has a puck-style groove of standard dimensions located in its base, [Fig. 4(a,b)]. The puck groove typically has an inner diameter of about 22.8mm and a height of greater than 2mm to accept the fork end.

For operations requiring a more flexible and precise movement, flag-style handles are used. They are the standard attachment for vacuum samples holders. Wobble sticks fitted with a gripper-jaw end[50] can grab onto a flag handle and move it within a vacuum chamber with linear and rotational movements, translating hand movements from outside the chamber. Flag handles are typically 4.2x4.2mm square protrusions with a thickness of 1mm. This STM has a flag-style handle on its front side, [Fig. 4(c,d)].

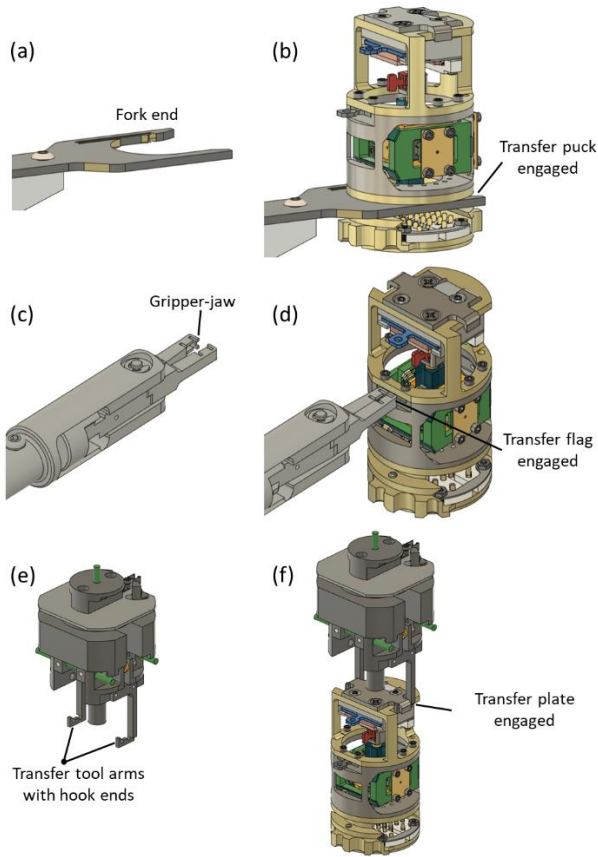

**FIG. 4:** Schematics showing transfer of STM head via three methods with vacuum compatible attachments. (a) Manipulator with fork end. (b) Fork end engaged with the puck in the base of the STM head. (c) Wobble stick manipulator with gripper-jaw end shown open. (d) Gripper-jaws closed around the flag handle on the front of the STM head. (e) Transfer tool equipped with two arms with hooked ends. This tool is a custom design used for vertical transfer. (f) Transfer tool arms engaged with the transfer plate on the top of the STM head.

The final transfer attachment on the STM head is custom made and used for vertically transferring the STM head into a top loading low temperature cryostat within UHV. A low-profile vertical transfer mechanism[51] was developed for this purpose. In this method, a manipulator tool is equipped with two arms with hooked ends. These hook arms can engage with the two handles of a transfer plate located on the top of the STM head, [Fig. 4(e,f)] enabling insertion into a connector plug located at the bottom of the cryostat. The details of this vertical transfer method will be published elsewhere.

## V. NESTED COARSE WALKER MODULE

For ease of transfer and stability, STM head size must be small enough to fit through load-locks and between vacuum ports, as well as within the bore of a magnet. This can be accomplished by designing a short and wide STM head which enhances stability and is especially important for connectors on vibration isolation stages where balance is a concern. The largest component of an STM head is the coarse walker module which controls the large-scale movement of the tip or sample. The primary role of a coarse walkers is to position the tip within nanometers of the sample surface (Z direction), in the region of interest. To prevent crashing the tip into the sample, precise and reliable control of the Z walker is required. To allow for sample and tip exchange, the overall Z travel must be large. The secondary action of the coarse walkers is to locate specific sample areas by controlling the X and Y travel. Large XY travel is crucial for locating μm-sized 2D samples as it allows navigation of the tip to the sample area[52].

Numerous coarse walker designs have been created to handle these requirements. The best known are the louse[13,14,22], the Beetle (Besocke)[53,54], and the Pan[18,55,56] types. The louse type was used in the first STM and allowed for both Z and X movement at RT and LT (4K). The Beetle type, created in 1987, has been used in both low and high temperature applications, but has limited X-Y movement control. The Pan type, developed in 1993, is widely used in cryogenic STM systems due to its rigid structural design providing high vibrational stability.

In the Pan arrangement, a single walker is composed of a sliding component, typically a sapphire prism, with three highly polished sides clamped inside a structure by six piezoelectric shear stacks, one pair of stacks on each side. Two of the piezo stack pairs are fixed within the mounting structure and the third pair is attached to a spring plate which is pressed against the sliding component with an adjustable tension. To make the sliding component walk, the piezo stacks are driven in a slip-stick pattern with a voltage waveform. The slip-stick pattern enables the piezo stacks to move together in one direction quickly, leaving the sliding component stationary (slip phase), and then slowly move the piezo stacks in the opposite direction, allowing the sliding component to move with them (stick phase).

The Pan-type walker provides reliable movement in a specified direction within a highly stable structure. Pan walkers in orthogonal directions can be stacked on top of each other to gain full range of motion. A standard configuration of an STM head mounts the tip to the Z walker and the sample to stacked X and Y walkers. Because each component is stacked in a linear fashion in this arrangement, the overall size of a standard STM head is large.

Researchers[37-40,57] have started to decrease the STM size by integrating the X and Y walkers into a single unit where the piezo stacks can move a sliding component in both the X and Y directions while it is clamped in a lateral plane. Our first version of the coarse walker utilized this integrated XY concept, and the size was reduced further by nesting the XY module within the Z module. This significantly reduced the STM head size, but we found that merging the XY piezo stacks caused the X and Y movement to become coupled together which made it difficult to control the motion. While this arrangement is sufficient for studying bulk sample where precise XY movement is not needed, it is not suitable for studying μm-sized samples. Consequently, it was necessary to develop a new concept based on the nested walker module, but with orthogonal drivers.

We created a compact coarse walker module comprised of the X, Y, and Z walkers nested inside each other while allowing for independent orthogonal travel over large distances in each direction, all contained within a compact unit [Fig. 5]. The X walker is the outermost unit built into the frame of the STM head. The Y walker is nested within the sliding component of the X walker, while the innermost Z walker is nested within the sliding component of the Y walker. The Z walker's sliding component is a sapphire prism that holds the piezoelectric scanner. Travel distances are X=8mm, Y=4.5mm, and Z=10mm, with nanometer step size. With this arrangement, the coarse motion only moves the tip, allowing the sample to have a stable fixed attachment to the STM head. This design also assists in tip and sample exchange by limiting the number of flexible and fragile components.

Each individual walker structure is based on the Pan-type design, described above. Titanium was used in the nested coarse walker module due to its high strength and thermal properties similar to BeCu, sapphire, alumina, and the piezoelectric material PZT. The X, Y, and Z walkers have similar piezo stack arrangements. The X walker [Fig. 5(c)] contains six piezo stacks built into the titanium STM head frame and used to clamp a sliding component. Four of the piezo stacks are fixed in place with insulating epoxy[43] to the frame and two are mounted to a spring-loaded titanium plate, used to clamp the sliding component. The spring-loaded plate tension is controlled through a ruby sphere[58] with a BeCu[59] spring sheet. The ruby sphere is used as a single point contact to the spring-loaded titanium plate to give the plate freedom to conform to the sliding component's surface. The BeCu spring sheet tension is adjusted by four titanium socket head screws, accessible on the back of the STM head. The sliding component of the X walker consists of three highly polished sapphire plates[60] mounted to the titanium body of the Y walker with insulating epoxy[43]. The X walker moves its sliding component side to side in the STM head. The Y walker adjustable spring plate screws are accessible on the side of the STM head [Fig. 5(c)]. The sliding component of the Y walker consists of three highly polished sapphire plates mounted to the titanium body of the Z walker. The Y walker moves its sliding component forward and backward in the STM head. Cut outs are made in the Y walker body to allow access to the adjustable spring plate of the Z walker. The Z walker spring plate screws are accessed via the front of the STM head [Fig. 5(c)]. The spring plates of all three walkers are compatible with *in-situ* vacuum adjustment. The Z walker sliding component is a sapphire prism[60]. Within the sapphire prism is encased the piezoelectric scanner[61] [Fig. 6(a)]. The piezoelectric scanner is used for safely bringing the tip into tunneling range of the sample and for fine movement control of the tip when scanning the sample surface. To mount and position the piezoelectric scanner within the sapphire prism, its end was fixed to a Macor insert which was mounted within the sapphire prism with insulating epoxy[43].

The piezo stacks for all walkers [Appendix Fig. 8] were constructed in-house. They are composed of two piezo shear plates[62] oppositely oriented to provide movement in a single direction with alumina wear plates[63] capping the top and bottom of the stack. BeCu sheets[64] are inserted between each layer for electrical contact. Conductive epoxy[47] fixes the layers of the stack together. A hole in the BeCu electrode between the stack layers provides better epoxy interaction between the layers and improved stack strength. Another hole in the electrode extending out of the stack is used for wire connections. Details of the stack assembly are provided in Appendix.

## VI. TIP AND SAMPLE HOUSING

The top portion of the STM head contains the tip and sample. This area has an open design to allow for optical access during tip and sample exchange and alignment. Tips with a diameter of 0.25mm are inserted in a capillary tube held in a flag-based tip holder, [Fig. 6(a)]. A flag-based structure is used for compatibility with vacuum manipulators for *in-situ* tip exchange. The tip holder can be plugged into a tip receiver in the STM head. The tip receiver has two springs[42] that fix the tip holder in place. A hole in the bottom of the tip receiver is used for electrically connecting the tunneling wire and for mounting to an insulating Macor piece. The Macor piece attaches to the top of the piezoelectric scanner with insulating epoxy[43].

The piezoelectric scanner is a fragile component that can easily break under excessive force. During tip exchange in a vacuum system, a structure support aid reduces force on the piezoelectric scanner, [Fig. 6(b)]. The support aid has a flag handle for manipulation in vacuum, hooks that can grab onto the top of the STM head to lock it in place, and an opening at its base to engage with the tip receiver to limit its movement. A typical tip exchange is carried out in a sequence of steps. First, the support aid is installed in the STM head. This is done by lowering the piezoelectric scanner, inserting the support aid into the STM head, lowering the support aid until its top hooks grab the top of the STM frame and lock it into position. Second, the coarse Z walker raises the piezoelectric scanner and tip receiver into the opening at the base of the support aid. The support aid holds the tip receiver and prevents it from moving in the XY plane. Third, the tip holder is inserted into the tip receiver while the receiver is supported. Fourth, the tip holder and receiver are lowered out of the support aid. Finally, the support aid is raised to disconnect its hooks from the STM top frame and remove it from the STM head, ending the tip exchange process.

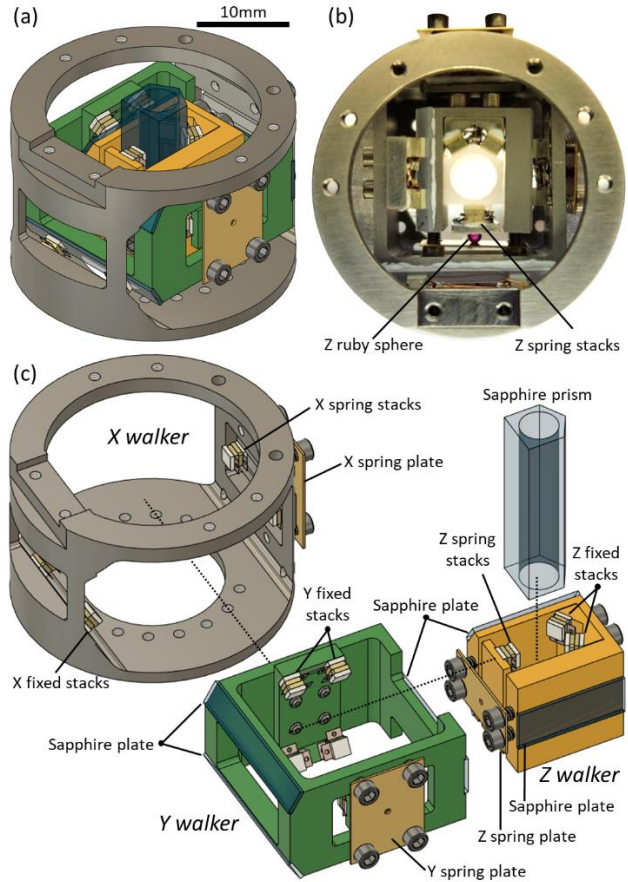

**FIG. 5:** Nested coarse walker module. (a) Schematic of nested coarse walker. X is the outermost walker (shown in grey), Y is the middle walker (green), and Z is the innermost walker (yellow). (b) Top view photograph of nested walker. The Z walker piezo stacks are shown engaged with the sapphire prism. (c) Exploded schematic view of nested coarse walker. The X, Y, and Z piezo stacks in the fixed and spring arrangements are indicated. The sapphire plates and prism are shown (blue).

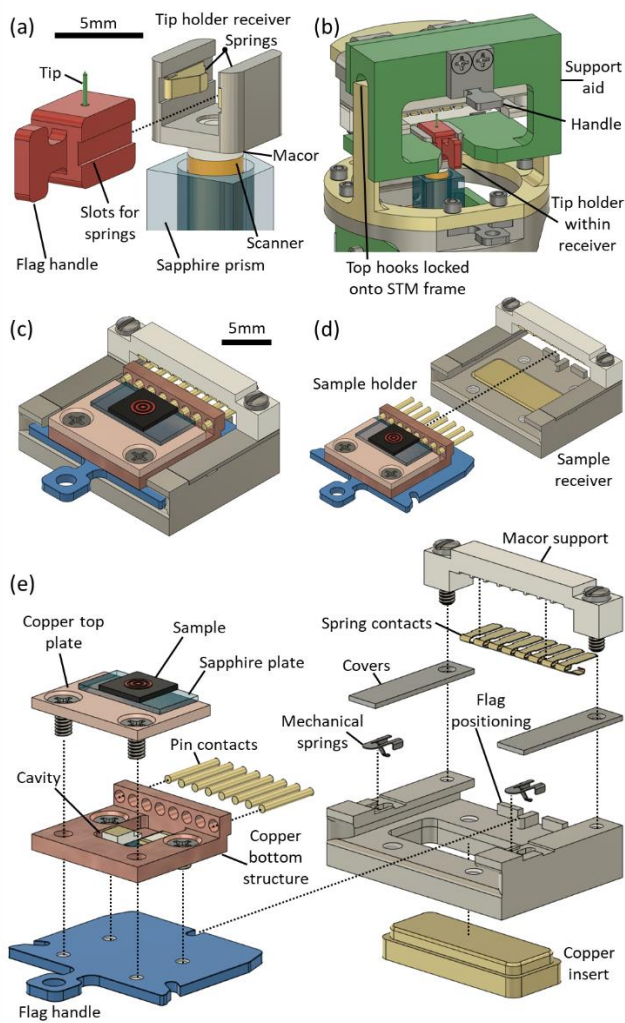

**FIG. 6:** Schematics of the tip and sample holders and receivers. (a) Flag tip holder with tip installed and tip receiver mounted to the piezoelectric scanner. Two springs hold the tip holder to the tip receiver. (b) Tip transfer support aid connected to STM head for tip exchange. Image depicts the tip receiver raised into the support aid and a tip holder inserted. (c) Sample holder installed within sample receiver. (d) Sample holder unplugged from the sample receiver. (e) Exploded-view of the sample holder and receiver.

Samples are mounted to a flag style sample holder and held within the STM head with a sample receiver, [Fig. 6(c,d)]. The STM provides a strong mechanical hold on samples while being electrically isolated from them. Sample substrates, typically $SiO_2$/Si of size 5x5mm, or bulk crystals are epoxied[47] to a sapphire plate which is mounted on a copper flag holder with electrical contacts, [Fig. 6(e)]. The sapphire plate is used to electrically isolate the sample from the flag while providing thermal conduction. The copper flag holder is composed of a top plate, bottom structure, and flag-style plate held together with screws. When the top plate is removed, a cavity within the bottom structure is accessible. The cavity is used for mounting a local heater and temperature sensor[65].

The flag is equipped with eight electrical pin[66] contacts which are attached to the bottom copper structure with insulating epoxy[43].

The sample flag plugs into the sample receiver via eight spring contacts[42] that are mounted[43] within an insulating Macor support near the back of the sample receiver, [Fig. 6(e)]. This connection and two additional mechanical springs[67], located along the sides of the receiver and held in place with covers, firmly fix the position of the sample flag. The sample receiver is made of titanium and has a guide structure near its back to assist flag positioning. The center of the receiver has a gold-plated copper insert to improve thermal contact between the sample flag and the frame. A temperature sensor[65] is mounted within the copper insert. Holes were punched in the sample receiver spring contacts to allow for simple wire attachment.

## VII. PERFORMANCE

This STM head was operated in a closed-cycle cryostat UHV system and a $^3$He system [Fig. 2]. These systems have vibration damping control[68] in place to minimize mechanical noise. The rigid construction of the STM head increases its resonant frequency, reducing interference from low frequency vibrations.

To demonstrate the high stability of the STM head, we tested its operation without vibration damping. In ambient conditions, the STM head was plugged into a connector mounted on a table with no vibration isolation or damping, shown in [Fig. 2]. Atomic resolution topography on the surface of an HOPG sample was obtained [Fig. 7(a)] with a PtIr tip using an RHK R9plus controller and an IVP-300 current preamplifier. Data was obtained for sample bias of 500mV and tunneling current of 200mA. A typical STS dI/dV spectrum taken at RT is shown in Fig. 7(b) with sample bias 200mV, current 200pA, and a 3.542kHz sinusoidal modulation voltage of 2mV. Further performance data of the STM head operating in UHV at near 4K and in the presence of a magnetic field within a closed-cycle cryostat will be published elsewhere.

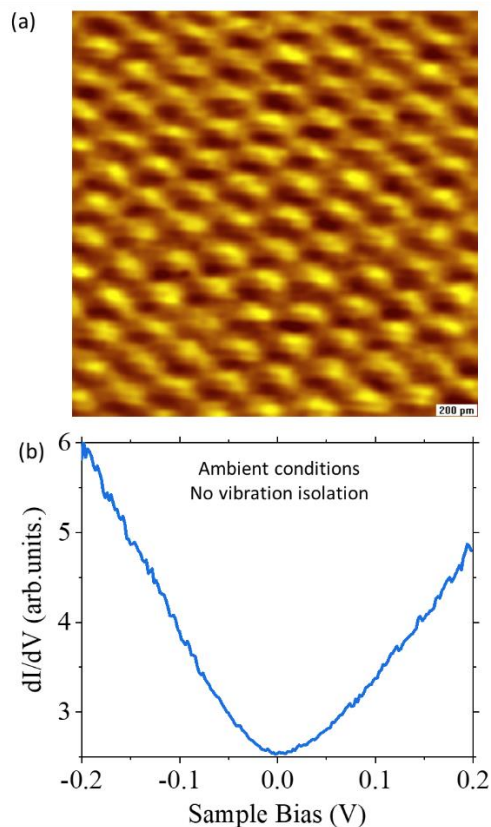

**FIG. 7:** STM measurements of HOPG under ambient conditions on a desk table with no vibration isolation. (a) High resolution topography STM image of HOPG showing the atomic structure. Image size is 2nm x 2nm. Tunneling parameter: tunneling current setpoint 200pA, sample bias 500mV. (b) Spectroscopic STS measurements of HOPG. dI/dV spectrum obtained with 200mV, 200pA, and 2mV modulation at 3.524kHz.

## VIII. SUMMARY AND OUTLOOK

This article describes the design, construction, performance of a detachable STM head equipped with a quick-connect socket matched to a universal connector plug, featuring nested coarse walkers, and vacuum transfer attachments. The transferable nature of the STM allows the head to be operated in a variety of environments, including ambient conditions, ultra-high vacuum, cryogenic temperatures, and high magnetic fields. The stability of the STM head was shown by achieving atomic resolution with no vibration isolation on a desktop in ambient conditions. This STM head was operated in a cryogen-free UHV system at variable temperatures ranging from RT down to 4K, and magnetic fields up to 9T as well as in a $^3$He system with temperatures down to 300mK. The modular structure of the STM head design enables access to a suite of exchangeable probes. To date, a total of three quick-connect STM heads have been created, each with different capabilities. Future probes could be designed that extend the STM concept presented here to a universal SPM platform by incorporating a suite of quick-connect modules, such as a tuning fork based atomic force microscope, a magnetic force microscope, or an electrical probe module.

## ACKNOWLEDGMENTS

The authors thank the following for helpful discussions and/or technical assistance: Paul Pickard, William Schneider, Eric Paduch, and Cole Woloszyn from the Rutgers Physics and Astronomy Machine Shop. The instrumentation and infrastructure were primarily supported by DOE-FG02-99ER45742, NSF-MRI 1337871, and Gordon and Betty Moore Foundation GBMF9453.

## AUTHOR DECLARATIONS:

A.M.C., G.L., and E.Y.A have Patent No. US 11,474,127 B2 issued. G.L. and E.Y.A have Patent No. WO 2019/209592 A2 pending. A.M.C., G.L., and E.Y.A have Patent No. WO 2021/226354 A1 pending.

## DATA AVAILABILITY:

The data that support the findings of this study are available from the corresponding author upon reasonable request.

## APPENDIX: Piezo stack assembly

Piezo stacks have specific requirements that need to be considered during assembly. The piezo stack's alumina wear plates and piezo shear plates need to be aligned together, [Fig. 8(a-c)]. To accomplish this, piezo stacks are assembled into a temporary corner structure that forms a 90 degree angle, [Fig. 8(d,e)]. Two sides of the piezo stack elements will align to the corner walls, forcing each element to be straight and aligned with each other. The BeCu electrodes stick out of the piezo stack on the sides opposite the corner walls. Two of the BeCu electrodes are ground lines which are kept physically separated from the voltage line by orienting them on different sides of the piezo stack. Electrically conductive epoxy[47] is placed in between each element in the stack. The top alumina wear plate of the stack must be kept pristine to interface with the sapphire sliding component. To accomplish this, the height of the temporary corner structure is made smaller than the height of the assembled piezo stack, removing the possibility of excess epoxy spreading to the top wear plate surface. Once the piezo stack was fully assembled, it was clamped to a fixed height and the corner structure was removed. Excess epoxy was removed from the piezo stack until its surfaces were pristine. Care was taken to prevent epoxy spread to the top alumina surface during the cleaning process. The piezo stacks were then cured in air (120C, 45min). The finished piezo stacks are mounted with insulating epoxy[43] to the titanium components. The piezo stack alignment to the sapphire sliding component was achieved by clamping the sapphire component between the piezo stacks before the adhesive was finished curing.

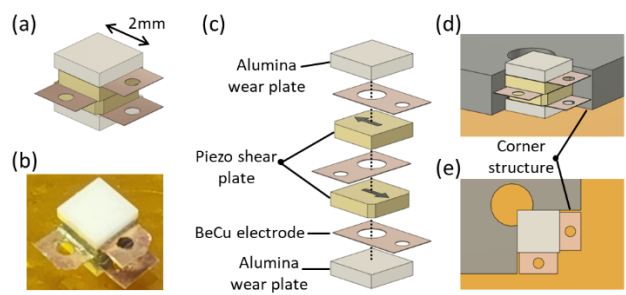

**FIG. 8:** Piezo stack assembly. (a) Schematic of assembled piezo stack. (b) Photograph of assembled stack. (c) Exploded schematic of piezo stack. Arrows on the piezo shear plates indicate the direction of polarization. Each BeCu sheet contains two holes for ease of assembly. (d) Side view and (e) top view schematic of piezo stack in corner structure during assembly process.

## REFERENCES:

1 Fischer, Ø., Kugler, M., Maggio-Aprile, I., Berthod, C. & Renner, C. Scanning tunneling spectroscopy of high-temperature superconductors. *Reviews of Modern Physics* **79**, 353-419, doi:10.1103/revmodphys.79.353 (2007).
2 Wiesendanger, R. Spin mapping at the nanoscale and atomic scale. *Reviews of Modern Physics* **81**, 1495-1550, doi:10.1103/revmodphys.81.1495 (2009).
3 Giessibl, F. J. Advances in atomic force microscopy. *Reviews of Modern Physics* **75**, 949-983, doi:10.1103/revmodphys.75.949 (2003).
4 Hla, S.-W. & Rieder, K.-H. STM Control of Chemical Reactions: Single-Molecule Synthesis. *Annual Review of Physical Chemistry* **54**, 307-330, doi:10.1146/annurev.physchem.54.011002.103852 (2003).
5 Bonnell, D. A. G., J. Scanning probe microscopy of oxide surfaces: atomic structure and properties. *Reports on Progress in Physics* **71**, doi:10.1088/0034-4885/71 (2008).
6 Kalinin, S. V. *et al.* Nanoscale Electromechanics of Ferroelectric and Biological Systems: A New Dimension in Scanning Probe Microscopy. *Annual Review of Materials Research* **37**, 189-238, doi:10.1146/annurev.matsci.37.052506.084323 (2007).
7 Schulte, A., Nebel, M. & Schuhmann, W. Scanning Electrochemical Microscopy in Neuroscience. *Annual Review of Analytical Chemistry* **3**, 299-318, doi:10.1146/annurev.anchem.111808.073651 (2010).
8 Kirtley, J. R. Fundamental studies of superconductors using scanning magnetic imaging. *Reports on Progress in Physics* **73**, doi:10.1088/0034-4885/73 (2010).

9. Huey, B. D. AFM and Acoustics: Fast, Quantitative Nanomechanical Mapping. *Annual Review of Materials Research* **37**, 351-385, doi:10.1146/annurev.matsci.37.052506.084331 (2007).
10. Novotny, L. & Stranick, S. J. Near-Field Optical Microscopy and Spectroscopy with Pointed Probes. *Annual Review of Physical Chemistry* **57**, 303-331, doi:10.1146/annurev.physchem.56.092503.141236 (2006).
11. Héctor, M. S. e. a. Hybrid strategies in nanolithography. *Reports on Progress in Physics* **73**, doi:10.1088/0034-4885/73 (2010).
12. van Houselt, A. & Zandvliet, H. J. W. Colloquium: Time-resolved scanning tunneling microscopy. *Reviews of Modern Physics* **82**, 1593-1605, doi:10.1103/revmodphys.82.1593 (2010).
13. Binnig, G., Rohrer, H., Gerber, C. & Weibel, E. Surface Studies by Scanning Tunneling Microscopy. *Physical Review Letters* **49**, 57-61, doi:10.1103/physrevlett.49.57 (1982).
14. Binnig, G., Rohrer, H., Gerber, C. & Weibel, E. Tunneling through a controllable vacuum gap. *Applied Physics Letters* **40**, 178-180, doi:10.1063/1.92999 (1982).
15. Binnig, G. & Rohrer, H. Scanning tunneling microscopy---from birth to adolescence. *Reviews of Modern Physics* **59**, 615-625, doi:10.1103/RevModPhys.59.615 (1987).
16. Tessmer, S. H., Harlingen, D. J. V. & Lyding, J. W. Integrated cryogenic scanning tunneling microscopy and sample preparation system. *Review of Scientific Instruments* **65**, 2855-2859, doi:10.1063/1.1144628 (1994).
17. Wildöer, J. W. G., van Roy, A. J. A., van Kempen, H. & Harmans, C. J. P. M. Low-temperature scanning tunneling microscope for use on artificially fabricated nanostructures. *Review of Scientific Instruments* **65**, 2849-2852, doi:10.1063/1.1144626 (1994).
18. Wittneven, C., Dombrowski, R., Pan, S. H. & Wiesendanger, R. A low-temperature ultrahigh-vacuum scanning tunneling microscope with rotatable magnetic field. *Review of Scientific Instruments* **68**, 3806-3810, doi:10.1063/1.1148031 (1997).
19. Ferris, J. H. *et al.* Design, operation, and housing of an ultrastable, low temperature, ultrahigh vacuum scanning tunneling microscope. *Review of Scientific Instruments* **69**, 2691-2695, doi:10.1063/1.1149000 (1998).
20. Foley, E. T., Kam, A. F. & Lyding, J. W. Cryogenic variable temperature ultrahigh vacuum scanning tunneling microscope. *Review of Scientific Instruments* **71**, 3428-3435, doi:10.1063/1.1287046 (2000).
21. Stipe, B. C., Rezaei, M. A. & Ho, W. A variable-temperature scanning tunneling microscope capable of single-molecule vibrational spectroscopy. *Review of Scientific Instruments* **70**, 137-143, doi:10.1063/1.1149555 (1999).
22. Eigler, D. M. & Schweizer, E. K. Positioning Single Atoms with a Scanning Tunnelling Microscope. *Nature* **344**, 524, doi:10.1038/344524a0 (1990).
23. de Lozanne, A. L., Elrod, S. A. & Quate, C. F. Spatial Variations in the Superconductivity of $Nb_3Sn$ Measured by Low-Temperature Tunneling Microscopy. *Physical Review Letters* **54**, 2433-2436, doi:10.1103/PhysRevLett.54.2433 (1985).
24. Li, G. & Andrei, E. Y. Observation of Landau levels of Dirac fermions in graphite. *Nature Physics* **3**, 623-627, doi:10.1038/nphys653 (2007).
25. Li, G., Luican, A. & Andrei, E. Y. Scanning Tunneling Spectroscopy of Graphene on Graphite. *Physical Review Letters* **102**, 176804, doi:10.1103/PhysRevLett.102.176804 (2009).
26. Luican, A., Li, G. & Andrei, E. Y. Quantized Landau level spectrum and its density dependence in graphene. *Physical Review B* **83**, 041405, doi:10.1103/PhysRevB.83.041405 (2011).
27. Kotta, E. *et al.* Spectromicroscopic measurement of surface and bulk band structure interplay in a disordered topological insulator. *Nature Physics* **16**, 285-289, doi:10.1038/s41567-019-0759-2 (2020).
28. Dai, J. *et al.* Restoring pristine $Bi_2Se_3$ surfaces with an effective Se decapping process. *Nano Research* **8**, 1222-1228, doi:10.1007/s12274-014-0607-8 (2015).
29. Koirala, N. *et al.* Record Surface State Mobility and Quantum Hall Effect in Topological Insulator Thin Films via Interface Engineering. *Nano Letters* **15**, 8245-8249, doi:10.1021/acs.nanolett.5b03770 (2015).
30. Meyer, G. A simple low-temperature ultrahigh-vacuum scanning tunneling microscope capable of atomic manipulation. *Review of Scientific Instruments* **67**, 2960-2965, doi:10.1063/1.1147080 (1996).
31. Pietzsch, O., Kubetzka, A., Haude, D., Bode, M. & Wiesendanger, R. A low-temperature ultrahigh vacuum scanning tunneling microscope with a split-coil magnet and a rotary motion stepper motor for high spatial resolution studies of surface magnetism. *Review of Scientific Instruments* **71**, 424-430, doi:10.1063/1.1150218 (2000).
32. Andrei, E. Y., Li, G. & Du, X. Electronic properties of graphene: a perspective from scanning tunneling microscopy and magnetotransport. *Reports on Progress in Physics* **75**, 056501, doi:10.1088/0034-4885/75/5/056501 (2012).
33. Altvater, M. A. *et al.* Charge Density Wave Vortex Lattice Observed in Graphene-Passivated $1T\text{-}TaS_2$ by Ambient Scanning Tunneling Microscopy. *Nano Letters* **21**, 6132-6138, doi:10.1021/acs.nanolett.1c01655 (2021).
34. Altvater, M. A. *et al.* Observation of a topological defect lattice in the charge density wave of 1T-TaS2. *Applied Physics Letters* **119**, doi:10.1063/5.0059662 (2021).
35. Lu, C.-P., Li, G., Mao, J., Wang, L.-M. & Andrei, E. Y. Bandgap, Mid-Gap States, and Gating Effects in $MoS_2$. *Nano Letters* **14**, 4628-4633, doi:10.1021/nl501659n (2014).
36. Lu, C.-P., Li, G., Watanabe, K., Taniguchi, T. & Andrei, E. Y. $MoS_2$: Choice Substrate for Accessing and Tuning the Electronic Properties of Graphene. *Physical Review Letters* **113**, doi:10.1103/physrevlett.113.156804 (2014).
37. Song, Y. J. *et al.* Invited Review Article: A 10 mK scanning probe microscopy facility. *Review of Scientific Instruments* **81**, doi:10.1063/1.3520482 (2010).
38. Celotta, R. J. *et al.* Invited Article: Autonomous assembly of atomically perfect nanostructures using a scanning tunneling microscope. *Review of Scientific Instruments* **85**, doi:10.1063/1.4902536 (2014).
39. Wong, D. *et al.* A modular ultra-high vacuum millikelvin scanning tunneling microscope. *Review of Scientific Instruments* **91**, doi:10.1063/1.5132872 (2020).
40. Schwenk, J. *et al.* Achieving µeV tunneling resolution in an in-operando scanning tunneling microscopy, atomic force microscopy, and magnetotransport system for quantum materials research. *Review of Scientific Instruments* **91**, doi:10.1063/5.0005320 (2020).
41. Coe, A. M., Li, G. & Andrei, E. Y. Patent. Pub. No.: US 11,474,127 B2. Application No.: 17/313,743. Modular Scanning Probe Microscope Head. (2021).
42. *Model 1447360-8, C-Clip PCB Spring Contact, Gold plated BeCu, TE Connectivity, Berwyn, PA.*
43. *Model 9530001, Torr Seal Low Vapor Pressure Resin Sealant, Agilent, Lexington, MA.*
44. *CeramTec North America LLC, Laurens, SC.*
45. *Model 100065-0073, 42 AWG Micro Coax, Molex, Wellington, Connecticut.*
46. *Model 110824, Silver plated copper wire Kapton insulated 34 AWG solid, Accu-Glass Products, Inc., Valencia, California.*
47. *Model H20E, Electrically conductive silver epoxy, Epoxy Technology, Billerica, MA.*
48. *Model RTTA 2-Axis, 2-Axis RTTA Rotary and Telescopic Extention with puck style end option, UHV Designs, East Sussex, UK.*
49. *Model WS40, Wobble stick with linear and rotary motion with puck style end option, UHV Designs, East Sussex, UK.*
50. *Model WSG40, Wobble stick with linear and rotary motion with gripper-jaw end option, UHV Designs, East Sussex, UK.*
51. Li, G. & Andrei, E. Y. Patent. Pub. No.: WO 2019/209592 A2. Application No.: PCT/US2019/027929. Low Profile Transfer Mechanism for Controlled Environments. (2019).
52. Li, G., Luican, A. & Andrei, E. Y. Self-navigation of a scanning tunneling microscope tip toward a micron-sized graphene sample. *Review of Scientific Instruments* **82**, doi:10.1063/1.3605664 (2011).
53. Besocke, K. An easily operable scanning tunneling microscope. *Surface Science* **181**, 145-153, doi:doi.org/10.1016/0039-6028(87)90151-8 (1987).
54. Frohn, J., Wolf, J. F., Besocke, K. & Teske, M. Coarse tip distance adjustment and positioner for a scanning tunneling microscope. *Review of Scientific Instruments* **60**, 1200-1201, doi:10.1063/1.1140287 (1989).
55. Pan, S. H., Hudson, E. W. & Davis, J. C. He3 refrigerator based very low temperature scanning tunneling microscope. *Review of Scientific Instruments* **70**, 1459-1463, doi:10.1063/1.1149605 (1999).
56. Pan, S. H. Patent. Pub. No.: WO 93/19494. Application No.: PCT/GB93/000539. Piezoelectric Motor. (1993).

57 Hug, H. J. *et al.* A low temperature ultrahigh vaccum scanning force microscope. *Review of Scientific Instruments* **70**, 3625-3640, doi:10.1063/1.1149970 (1999).
58 *Model AJ60-SP-000130, Ruby Balls 1.5mm diameter, Goodfellow Corporation - USA, Pittsburgh, PA.*
59 *Model CU07-FL-000151, Copper/Beryllium Alloy Foil 0.25mm thick half hard 300mm coil W, Goodfellow Corporation - USA, Pittsburgh, PA.*
60 *INSANCO Inc., Quakertown, PA.*
61 *Model EBL#4, Piezoceramic tube 0.125"OD x 0.020" wall x 0.500" length with 90deg electrode quadrants on OD to legth midpoint and single electrode on remaining OD and ID, EBL Producs Inc., East Harford, CT.*
62 *Model Noliac CSAP01, Shear actuator 2x2x0.5mm, Micromechtronics, State College, PA.*
63 *Model White alumina 96% wear plates 2x2x0.5mm, Micromechtronics, State College, PA.*
64 *Model CU07-FL-000175, Copper/Beryllium Alloy Foil 2x1.8x0.02mm, Goodfellow Corporation - USA, Pittsburgh, PA.*
65 *Model Cernox 1050-SD-HT-P, Temperature range 2-420K, Lakeshore Cryotronics, Woburn, MA.*
66 *Model PCB 3330-0-00-15-00-00-03, Circuit Board hardware, Mill-Max, Oyster Bay, NY.*
67 *Model 2199248-6, C-Clip PCB Spring Contact, TE Connectivity, Berwyn, PA.*
68 Coe, A. M., Li, G. & Andrei, E. Y. Patent. Pub. No.: WO 2021/226354 A1. Application No.: PCT/US2021/031120. Damping Base for Modular Scanning Probe Microscope Head. (2021).